\documentclass[aps, longbibliography, amsmath,amssymb, superscriptaddress, 12pt]{revtex4-1}

\usepackage{graphicx}
\usepackage{dcolumn}
\usepackage{bm}
\usepackage{lineno}

\newcommand{\feses}{$\textrm{Fe}\textrm{Se}_{1-x}\textrm{S}_{x}$}

\newcommand{\fesesonenine}{$\textrm{Fe}\textrm{Se}_{0.81}\textrm{S}_{0.19}$}

\newcommand{\fese}{$\textrm{Fe}\textrm{Se}$}

\newcommand{\gqe}{$g(\vec{q},E)$}

\newcommand{\bafeas}{$\textrm{Ba}\textrm{Fe}_{2}\textrm{As}_2$}

\usepackage{color,graphicx,subfigure}
\usepackage{enumitem}
\usepackage{wrapfig}
\usepackage{soul}
\usepackage{amsmath}
\usepackage{titlesec}
\usepackage{multirow}
\usepackage{textcomp}
\usepackage{gensymb}
\usepackage[]{hyperref}
\hypersetup{colorlinks=true,linkcolor=blue,citecolor=blue,urlcolor=blue,pdfpagemode=UseNone}
\usepackage{xcolor}
\usepackage{etoolbox}
\usepackage{floatrow}
\usepackage{floatflt}
\usepackage{datetime}

\definecolor{eh}{rgb}{1, 0, 0}
\definecolor{mw}{rgb}{0, .8, 0}
\definecolor{ag}{rgb}{0, 0, .8}

\begin{document}


\title{Superconductivity Mediated by Nematic Fluctuations in Tetragonal \feses}

\author{\small Pranab Kumar Nag}
\email[These authors contributed equally to this work.]{}
\affiliation{\footnotesize \mbox{Department of Physics, Yale University, New Haven, Connecticut 06520, USA}}
\affiliation{\footnotesize \mbox{Energy Sciences Institute, Yale University, West Haven, Connecticut 06516, USA}}

\author{\small Kirsty Scott}
\email[These authors contributed equally to this work.]{}
\affiliation{\footnotesize \mbox{Department of Physics, Yale University, New Haven, Connecticut 06520, USA}}
\affiliation{\footnotesize \mbox{Energy Sciences Institute, Yale University, West Haven, Connecticut 06516, USA}}

\author{\small Vanuildo S.\,de Carvalho}
\affiliation{\footnotesize \mbox{Instituto de F\'isica, Universidade Federal de Goi\'as, 74.001-970, Goi\^ania-GO, Brazil}}

\author{\small Journey K.\,Byland}
\affiliation{\footnotesize \mbox{Department of Physics and Astronomy, University of California, Davis, California 95616, USA}}

\author{\small Xinze Yang}
\affiliation{\footnotesize \mbox{Department of Physics, Yale University, New Haven, Connecticut 06520, USA}}
\affiliation{\footnotesize \mbox{Energy Sciences Institute, Yale University, West Haven, Connecticut 06516, USA}}

\author{\small Morgan Walker}
\affiliation{\footnotesize \mbox{Department of Physics, Yale University, New Haven, Connecticut 06520, USA}}
\affiliation{\footnotesize \mbox{Energy Sciences Institute, Yale University, West Haven, Connecticut 06516, USA}}
\affiliation{\footnotesize \mbox{Department of Physics and Astronomy, University of California, Davis, California 95616, USA}}

\author{\small Aaron~G.\,Greenberg}
\affiliation{\footnotesize \mbox{Department of Physics, Yale University, New Haven, Connecticut 06520, USA}}
\affiliation{\footnotesize \mbox{Energy Sciences Institute, Yale University, West Haven, Connecticut 06516, USA}}

\author{\small Peter Klavins}
\affiliation{\footnotesize \mbox{Department of Physics and Astronomy, University of California, Davis, California 95616, USA}}

\author{\small Eduardo Miranda}
\affiliation{\footnotesize \mbox{Gleb Wataghin Institute of Physics, University of Campinas, Campinas, S\~ao Paulo 13083-950, Brazil}}

\author{\small Adrian Gozar}
\affiliation{\footnotesize \mbox{Department of Physics, Yale University, New Haven, Connecticut 06520, USA}}
\affiliation{\footnotesize \mbox{Energy Sciences Institute, Yale University, West Haven, Connecticut 06516, USA}}

\author{\small Valentin Taufour}
\affiliation{\footnotesize \mbox{Department of Physics and Astronomy, University of California, Davis, California 95616, USA}}

\author{\small Rafael~M.\,Fernandes}
\affiliation{\footnotesize \mbox{School of Physics and Astronomy, University of Minnesota, Minneapolis, MN 55455, USA}}

\author{\small Eduardo H.\,da Silva Neto}
\email[Corresponding Author: ]{eduardo.dasilvaneto@yale.edu}
\affiliation{\footnotesize \mbox{Department of Physics, Yale University, New Haven, Connecticut 06520, USA}}
\affiliation{\footnotesize \mbox{Energy Sciences Institute, Yale University, West Haven, Connecticut 06516, USA}}
\affiliation{\footnotesize \mbox{Department of Physics and Astronomy, University of California, Davis, California 95616, USA}}
\affiliation{\footnotesize \mbox{Department of Applied Physics, Yale University, New Haven, Connecticut 06520, USA}}



\maketitle


\textbf{
Nematic phases, where electrons in a solid spontaneously break rotational symmetry while preserving the translational symmetry, exist in several families of unconventional superconductors \cite{Fradkin-ARCMP(2010),fernandes2014drives}.
Although superconductivity mediated by nematic fluctuations is well established theoretically \cite{Lederer_enhancement_PRL_2015,Metlitski2015,Lederer2017,Klein-PRB(2018),Kang_Fernandes2017}, it has yet to be unambiguously identified experimentally \cite{fernandes2022iron, Bohmer2022}. A major challenge is that nematicity is often intertwined with other degrees of freedom, such as magnetism and charge order. The \feses{ }family of iron based superconductors provides a unique opportunity to explore this concept, as it features an isolated nematic phase that can be suppressed by sulfur substitution at a quantum critical point (QCP) near $x_c=0.17$, where nematic fluctuations are the largest \cite{coldea2020electronic, hosoi2016nematic,Ayres2022transport}. 
Here, we performed scanning tunneling spectroscopy measurements to visualize Boguliubov quasiparticle interference patterns, from which we determined the momentum structure of the superconducting gap near the Brillouin zone $\Gamma$ point of \fesesonenine. The results reveal an anisotropic, near nodal gap with minima that are $45^\circ$ rotated with respect to the Fe-Fe direction, characteristic of a nematic pairing interaction, {contrary to the usual isotropic gaps due to spin mediated pairing in other tetragonal Fe-based superconductors.} 
{The results are also in contrast with pristine \fese{}, where the pairing is mediated by spin fluctuations and the gap minima are aligned with the Fe-Fe direction.}
Therefore, the measured gap structure demonstrates not only a fundamental change of the pairing mechanism across the phase diagram of \feses{}, but it also indicates the existence of superconductivity mediated by nematic fluctuations in \fesesonenine.    
}  

\vspace{5mm}

Having one of the clearest realizations of nematicity amongst unconventional superconductors, iron-based superconductors (FeSCs) are the most promising materials to search for superconductivity mediated by nematic fluctuations \cite{Bohmer2022}.
However, in most FeSCs the nematic phase appears in tandem with spin density wave (SDW) order, whose fluctuations have been argued to be the dominant interaction mediating Cooper pairing \cite{fernandes2022iron}.  
This is seen, for example, in the archetypal \bafeas{ }system, where superconductivity is most robust when the concomitant magnetic and nematic phases are suppressed by doping or pressure \cite{Shibauchi_ARCMP_2014, Hayes2016}. 
{Near such quantum critical points (QCPs), either spin or nematic fluctuations could theoretically serve as the primary pairing interaction promoting superconductivity, yet spin fluctuations prevail, yielding nearly isotropic superconducting gaps in momentum space.}
{Nevertheless, even in the presence of dominant spin-mediated pairing, nematic fluctuations may still participate in the pairing mechanism, potentially even enhancing the superconducting transition temperature ($T_c$).
Thus, elucidating the relationship between nematic fluctuations and superconductivity holds great importance, especially for comprehending unconventional superconductors where nematicity  exists, including not only FeSCs but also high-Tc cuprates and twisted bilayer graphene \cite{keimer2015quantum, Andrei2020}.}

\begin{figure}
    \centering
    \includegraphics[width=0.66\linewidth]{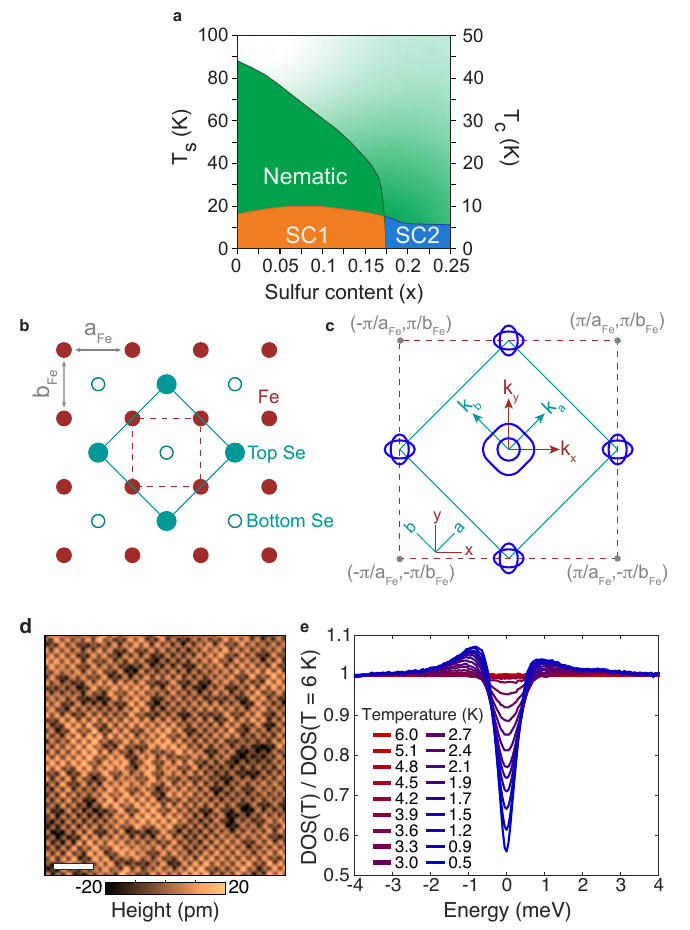}
    \caption{
    \footnotesize
    \textbf{Phase diagram of \feses, and electronic structure of superconducting \fesesonenine.}
    (\textbf{a}) Phase diagram of \feses{ }based on \cite{coldea2020electronic}, indicating the nematic/structural (tetragonal to orthorhombic) and superconducting transitions, $T_S$ and $T_c$. For better visualization, we show the $T_S$ and $T_c$ values on separate scales, left and right vertical axes, respectively. (\textbf{b}) Schematic top view of the crystal structure of tetragonal \feses{}. S atoms are expected to be randomly located on Se locations, but are omitted for clarity. Dashed lines represent the 1-Fe unit cell and solid lines show the actual crystallographic 2-Fe unit cell. (\textbf{c}) Schematic of the Fermi surface of tetragonal \feses. The sizes of the pockets were enlarged by a factor of two in momentum space. (\textbf{d}) Representative constant current topographic images of \fesesonenine{ }showing the atomically resolved (Se,S) termination layer. The S atoms are seen as cross like features in the data. The white scale bar represents $20$\,\AA.  (\textbf{e}) Spatially averaged DOS of \fesesonenine{ } showing the suppression of the superconducting gap with temperature.}
    \label{fig:1}
\end{figure}

Unfortunately, {the existence of so closely coupled SDW and nematic phases in most FeSCs hinders} our ability to disentangle the relation between nematic fluctuations and superconductivity, despite the strong evidence for nematic quantum criticality in compounds such as doped \bafeas{ }\cite{worasaran2021nematic}.
A notable exception occurs in the \feses{ }system, where the nematic phase is decoupled from magnetism \cite{coldea2020electronic, hosoi2016nematic, Ayres2022transport}. The substitution of S for Se suppresses the nematic phase towards a putative QCP at $x=x_c\approx0.17$ \cite{Licciardello2019,huang2020non,zhang2021quadrupolar,chibani2021lattice}, Fig\,\ref{fig:1}a, whereas the SDW phase is absent and appears only upon the application of pressure \cite{coldea2020electronic, hosoi2016nematic, matsuura2017maximizing}. In contrast to the enhancement of the superconducting transition temperature ($T_c$) at the putative SDW QCP in the doped \bafeas{ }compounds, there is no obvious enhancement of superconductivity near $x_c$ in \feses. While at first sight this could be interpreted as evidence against pairing mediated by nematic fluctuations, it can also be a consequence of a large nemato-elastic coupling \cite{Paul2017} or of long-range nematic order cooperating rather than competing with superconductivity for $x < x_c$ \cite{Hirschfeld2020}. {Interestingly, the maximum $T_c$ in \feses{} is observed deep inside the nematic state, near $x=0.1$ \cite{Ishida_PNAS_2022,Mukasa_PRX_2023}, where spin fluctuations are also the largest \cite{Wiecki_NMR_2018}. The momentum space ($\vec{k}$) structure of the gap in \fese{}, obtained from detailed scanning tunneling microscopy and spectroscopy (STM/S) measurements, is also consistent with superconductivity mediated by spin fluctuations \cite{Sprau_Science_orb_selective_cooper,Kang_Fernandes_Chubukov_PRL,Benfatto2018}. However, }across $x_c$, tunneling and thermodynamic measurements suggest significant changes of the superconducting gap and of the electronic structure \cite{Hanaguri_Sci_Adv, Sato_abrupt_2018,Coldea_npj_2019}, implying that different pairing mechanisms may be active inside (SC1) and outside (SC2) the long-ranged nematic state. {Thus, while there is abundant evidence for spin-mediated pairing in the SC1 phase, the pairing interaction in the SC2 phase, near the nematic QCP, remains unknown.}

{Despite potentially holding the key to unraveling the existence of superconductivity mediated by nematic fluctuations,} the gap function $\Delta(\vec{k})$ for $x\gtrsim x_c$, which is also the region where nematic fluctuations are most intense \cite{hosoi2016nematic}, has not yet been determined. 
One of the difficulties stems from the fact that, {compared to the average gap observed in \fese{ }($\Delta_{avr} \approx \pm 2.5$\,meV) and in typical pnictide FeSCs ($\Delta_{avr} \approx \pm 4$ to $12$\,meV), the average gap in the SC2 region is small ($\Delta_{avr} \approx \pm 0.8$\,meV)}, which has precluded the {spectroscopic} determination of $\Delta(\vec{k})$ in SC2 {by previous studies} \cite{Hanaguri_Sci_Adv}. In this work, we used subkelvin scanning tunneling microscopy and spectroscopy (STM/S) to measure the momentum structure of the superconducting gap on \fesesonenine{ }with high energy resolution. 
{Our main finding is that the gap structure for \feses{ }near the nematic QCP is highly anisotropic with deep minima along a direction $45^\circ$ from the Fe-Fe directions. This is in stark contrast to all other tetragonal FeSCs, where the gaps on the hole pockets are nearly isotropic or show shallow minima along the Fe-Fe direction, or both \cite{Richard_2015,Allan_LiFeAs}.
Remarkably, the observed gap structure in general contradicts theoretical predictions based on the spin-fluctuation scenario \cite{Graser_2009, Aoki_2009, Maier_2009, Ikeda_2010, Maiti_2011, yin2014spin, Rhodes2021, Fernandez_Martin_RPA}, yet it aligns with predictions for superconductivity mediated by nematic fluctuations \cite{Lederer_enhancement_PRL_2015,Klein-PRB(2018)}.}

The crystal structure of \feses{ }is either orthorhombic or tetragonal, inside or outside the nematic phase, respectively. Both 1-Fe (dashed squares) and 2-Fe (solid squares) unit cells (Fig.\ref{fig:1}b) and corresponding Brillouin zones (Fig.\ref{fig:1}c) are commonly used in the literature.  In the nematic phase, the two orthogonal Fe-Fe bonds become nonequivalent, $a_{Fe} \neq b_{Fe}$. Here we denote the directions along the Fe-Fe directions by $x$ and $y$, and refer to the crystal axes of the 2-Fe unit cell as $a$ and $b$, as shown in Fig.\,\ref{fig:1}c. The Fermi surface of tetragonal \feses{} consists of two hole-like bands surrounding the $\Gamma$ point and two electron-like bands at the M points of the 2-Fe unit cell \cite{coldea2020electronic}. To investigate the influence of nematic fluctuations on the superconductivity in SC2, we focused our studies on \fesesonenine{}, which is very close to the putative nematic QCP. A representative STM topographic image of the Se termination is shown in Fig.\,\ref{fig:1}d, where the S atoms appear as densely distributed cross-like features at the Se locations. Temperature dependent spectroscopic measurements reveal the suppression of the spatially averaged superconducting gap with increasing temperature, Fig.\,\ref{fig:1}e.

\begin{figure}
    \centering
    \includegraphics[width=6.5in]{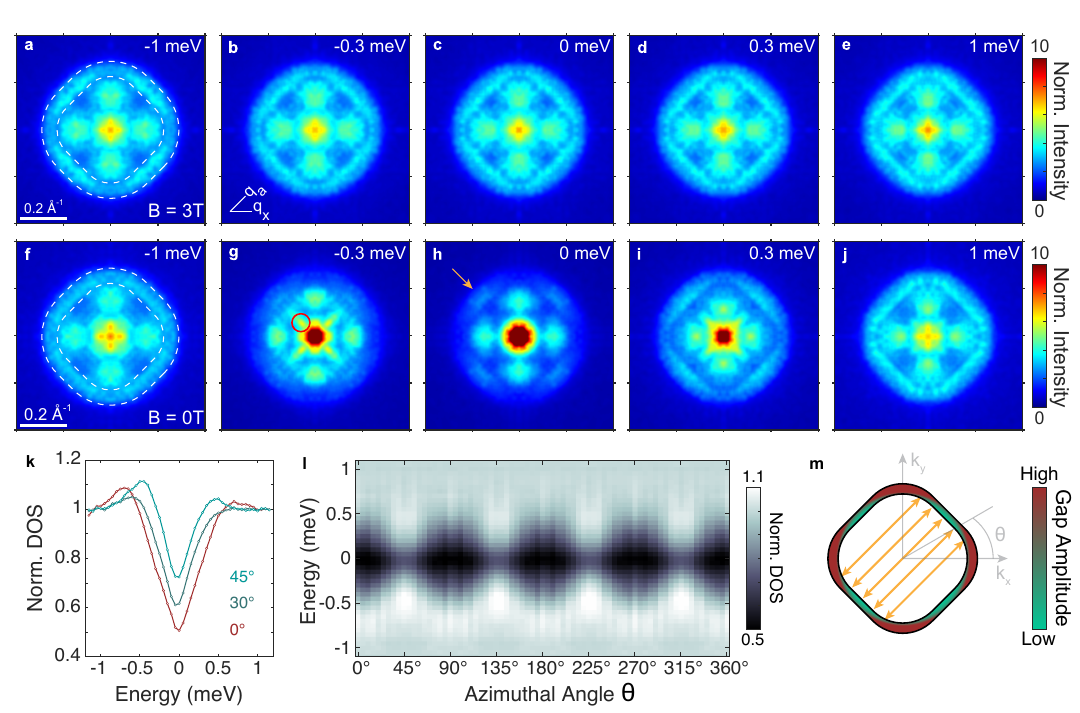}
    \caption{
    \footnotesize
    \textbf{QPI from the superconducting state in \fesesonenine{ }and identification of the gap anisotropy.}
    (\textbf{a-e}) Measured \gqe{ }at different energies $E$ in the non-superconducting state at $B=3$\,T. (\textbf{f-j}) Similar to (a-e) but for \gqe{ }measured in the superconducting state $B=0$\,T. (\textbf{k}) DOS from the rounded square QPI pattern (e.g. the area between the dashed lines in (f)) along different $\theta$, revealing an anisotropic gap structure. 
    (\textbf{l}) DOS map as a function of energy and $\theta$ from the rounded square pattern, clearly showing gap minima along $45^\circ$, $135^\circ$, $225^\circ$ and $315^\circ$.
    (\textbf{m}) Schematic of an anisotropic gap on a rounded square Fermi surface. The gap minima occur at the directions indicated by (l). Orange arrows indicate various scattering processes contributing to the intensity on the square pattern along $\theta=0^\circ$.  
    }
    \label{fig:2}
\end{figure}

\noindent \textbf{Direction of the Superconducting Gap Minima}\\
To probe the momentum structure of the superconducting gap we used STS to measure quasiparticle interference (QPI) patterns in \fesesonenine{}. 
QPI appears as periodic modulations of the density of states \cite{hoffman2002imaging}, with characteristic energy-dependent $\Vec{q}(E)$ wave vectors that reflect constant-energy scattering processes between different initial $\vec{k}_i$ and final $\vec{k}_f$ momentum states.
The Fourier transform of the STS images, $g(\vec{q},E)$, results in patterns that have been widely used to image the normal and superconducting band structures, nematicity and stripe-like states in \feses{ }\cite{Sprau_Science_orb_selective_cooper, Hanaguri_Sci_Adv, Watashige_FeSe_Domains, Moore_STM_2015_PRB, Walker_Stripes_2023}. 
Figures \ref{fig:2}a-e show \gqe{ }at a magnetic field $B = 3$\,T in the non-superconducting state. For $E=-1.0$\,meV, a clear rounded square pattern is observed (area between the two dashed white lines in Fig.\,\ref{fig:2}a), which decreases in size towards $E=+1$\,meV (Fig.\,\ref{fig:2}e). Its shape and hole-like dispersion are consistent with intra-band scattering originating from the outer pocket at $k_z = 0$ ($\Gamma$ point), as described in the Supplementary Note I.
In the superconducting state ($B=0$\,T), \gqe{ }patterns (Figs.\,\ref{fig:2}f-j) are significantly different for energies inside the gap. In the following, we divide our discussion of the \gqe{ }data into two regions: region I, the rounded-square contour area, and region II, the area of smaller momenta (\textit{e.g.}, $\vec{q}$ inside the inner white dashed square in Fig.\,\ref{fig:2}a). 
In region II, which we discuss in detail later, superconductivity induces new peaks, such as the one marked by the red circle in Fig.\,\ref{fig:2}g, which originate from Bogoliubov QPI (BQPI). In region I, the relative intensity between points along the $a$ and $x$ directions changes from nearly equal at $E=\pm1$\,meV (Figs.\,\ref{fig:2}f,j) to the intensity along $a$ dominating over the intensity along $x$ near $E=0$ (Figs.\,\ref{fig:2}g-i), as indicated by the orange arrow in Fig.\,\ref{fig:2}h. 
This observation prompted us to analyze the energy dependent DOS of the rounded square QPI feature (\textit{e.g.}, the area between dashed lines in Fig.\,2a) for different angles $\theta$, where $\theta$ is defined as the angle from $k_x$ towards $k_a$. Under this definition, the gap on the $\Gamma$ pocket of \fese{}, which is two-fold symmetric, has minima along $\theta = 90^\circ$ and $270^\circ$, as shown in Refs.\,\cite{Sprau_Science_orb_selective_cooper,Liu2018}. In contrast, for \fesesonenine, our analysis reveals a smaller gap at $\theta = 45^\circ$ than at $\theta = 0^\circ$, Figs.\,\ref{fig:2}k,l. In other words, while the gap minima in \fese{ }occur along the $y$ direction (or $x$, depending on the nematic domain), our measurements indicate that the gap minima in \fesesonenine{ }occur along the $a$ and $b$ axes. At first, one might be tempted to interpret the finite size of the QPI-extracted gap at $45^\circ$ as evidence for the absence of a node. However, the spectral intensity at $45^\circ$ on the square contour of \gqe{ }involves contributions from various $\vec{k}$ points on the Fermi surface, which have varying gap sizes, as depicted schematically in Fig.\,\ref{fig:2}m. Therefore, the location of the coherence peaks for the $45^\circ$ gap (approximately $\pm450$\,$\mu$eV) represents an upper bound for the actual value of the gap along $a$ or $b$, leaving open the possibility of a nodal structure. Regardless, the $\theta$ dependence of the spectra shown in Figs.\,\ref{fig:2}k,l already indicates a fundamental change in gap structure between the SC1 and SC2 phases. 

\begin{figure}
    \centering
    \includegraphics[width=\linewidth]{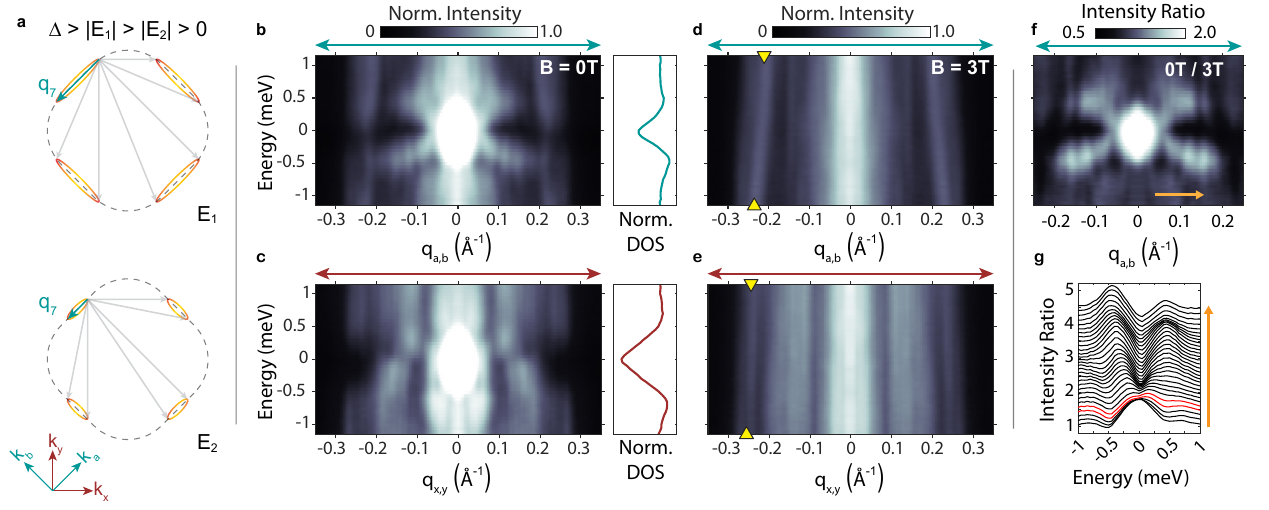}
    \caption{
    \footnotesize
    \textbf{Bogoliubov QPI in \fesesonenine.}
    (\textbf{a}) Schematic representation of constant-energy contours in the superconducting state at two different energies. The red color at the tips of the contours represent the larger DOS at those points. 
    (\textbf{b}-\textbf{e}) Dispersion plots measured at $B=0$\,T (b,c) and $B=3$\,T (d,e), along the $a$,$b$ direction (b,d) and the $x$,$y$ direction (c,e), as defined in Fig.\,\ref{fig:1}. The side panels in (b,c) are the gaps on the square contours of \gqe{}, also shown in Fig.\,\ref{fig:2}k. The yellow triangles delimit the dispersion of the rounded square feature shown in Fig.\,\ref{fig:2}a-e. 
    (\textbf{f}) Dispersion of the BPQI wave vector $\vec{q}_7$ obtained from the division of the $B=0$\,T data (b) by the $B=3$\,T data (d). 
    (\textbf{g}) Constant $\vec{q}$ cuts of the data in (f), vertically shifted for clarity. The range of momenta shown in (g) is from $0.0384$\,\AA$^{-1}$ to $0.1557$\,\AA$^{-1}$, orange arrow in (f), {equally spaced in steps of $0.0040$\,\AA$^{-1}$.} The red curves in (g) are for the lowest momenta ($0.0505$\,\AA$^{-1}$ and $0.0546$\,\AA$^{-1}$) where a two-peak structure is clearly observed.  
    }
    \label{fig:3}
\end{figure}

The direction of the gap minima in \fesesonenine{ }can also be determined via analysis of the BQPI signal in region II.
When a superconducting gap is anisotropic in $\vec{k}$ space, the Bogoliubov quasiparticle momentum structure exhibits closed constant-energy contours (CECs) that are distinct from their normal state counterparts. For a given $E$, these CECs always feature two points of largest DOS that are anchored to the underlying Fermi surface where $|\Delta(\vec{k})| = |E|$. For a rounded-square Fermi surface such as the one inferred from the \gqe{ }maps, the CECs are expected to promote seven BQPI wave vectors $\Vec{q}_i$ that connect eight points of high DOS, similar to the celebrated octet model for the cuprates \cite{hoffman2002imaging, wang2003quasiparticle,mcelroy2003octet}.
Among the seven wave vectors, $\vec{q}_7$, depicted by the blue arrow in Fig.\,\ref{fig:3}a, is especially important for identifying the directions of the gap minima. For instance, in the case of a nodal gap, $\vec{q}_7$ indicates the direction of the gap minima and it is the only wave vector that decreases to zero length at $E=0$. 
To identify the direction of $\vec{q}_7$ we compare \gqe{ }along the two high symmetry directions, $a$ and $x$ (Figs.\,\ref{fig:3}b-c) to their non-superconducting counterparts (Figs.\,\ref{fig:3}d-e). These dispersion plots show various superconductivity-induced modifications to \gqe{ }within the gap energies. The outermost QPI feature (\textit{i.e.} largest $|\Vec{q}|$, indicated by yellow triangles in Figs.\,\ref{fig:3}d-e) in the non-superconducting state corresponds to the rounded-square Fermi surface feature while peaks observed at smaller $|\vec{q}|$ likely originate from other hole or electron bands (Figs.\,\ref{fig:3}d-e). 
The most salient feature in these plots is the emergence of a BQPI feature along $q_{a,b}$ that decreases towards $|\Vec{q}| = 0$ at $E = 0$ and is approximately symmetric across the Fermi level, Fig.\,\ref{fig:3}b. In contrast, no such feature exists along the $q_{x,y}$ direction, Fig.\,\ref{fig:3}c. This X pattern in the dispersion, which has been confirmed over multiple measurements on different samples and under different experimental conditions (Supplementary Note II), clearly identifies $a$ and $b$ (i.e. $\theta = 45^\circ$) as the directions of smallest gap, consistent with the analysis shown in Fig.\,\ref{fig:2}.

\vspace{-0mm}
\begin{figure}
    \centering
    \includegraphics[width=1\linewidth]{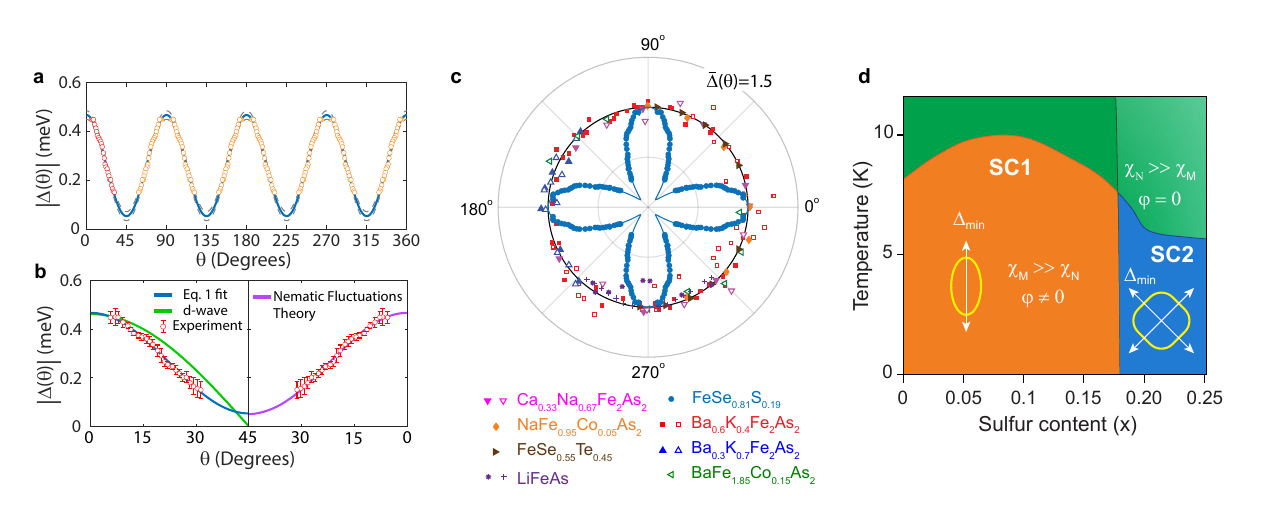}
    \caption{
    \footnotesize
    \textbf{Angular dependence of the superconducting gap in \fesesonenine.}
    (\textbf{a}) $|\Delta(\theta)|$ obtained from BQPI data (see Supplementary Note V for details). The red points between $0^\circ$ and $45^\circ$ are repeated for all other equivalent angles (orange). Blue line is a least squares fit of Eq.\,\ref{Eq:1} to the data, yielding ${\Delta^\prime_s} = 0.42 \pm 0.02$\,meV and $\Delta_s = 0.05 \pm 0.01$\,meV, where the uncertainties, as well as the gray dashed lines, are obtained from the $95$\% confidence intervals from the fit.
    (\textbf{b}) $|\Delta(\theta)|$ data in (a), with error bars representing $95$\% confidence intervals obtained from fitting Gaussian functions plus a linear background to the data in (b). Also shown are the Eq.\,\ref{Eq:1} fit (blue, same as in (d)), d-wave function fit $\Delta(\theta) = \Delta_d \cos(2\theta)$ (green, fit to the three lowest $\theta$ values, see Supplementary Note VI) and the theoretical calculation considering pairing due to nematic quantum critical fluctuations (see Supplementary Note V). 
    {(\textbf{c}) Polar plots comparing $|\bar{\Delta}(\theta)|$, the normalized gap, in \fesesonenine{} (outer hole pocket, full circles) and in various tetragonal FeSCs (reproduced from \cite{Richard_2015,Allan_LiFeAs}). 
    The gaps for the largest (stars \cite{Richard_2015}) and middle (crosses \cite{Allan_LiFeAs}) pockets at $\Gamma$ in LiFeAs, which has three pockets, are depicted, with the reported anisotropic gaps normalized to their respective maxima.
    For the other materials, open (full) symbols represent $|\bar{\Delta}(\theta)|$ for outer (inner) hole pockets. See Supplementary Note VII for the data without normalization.} 
    (\textbf{f}) Phase diagram of \feses{ }with the white arrows depicting the direction of $\Delta_{\mathrm{min}}$ on the hole Fermi surfaces in SC1 and SC2, and the presence or absence of magnetic fluctuations ($\chi_M$), nematic order ($\varphi$) and nematic fluctuations ($\chi_N$), as discussed in the text. 
    }
    \label{fig:4}
\end{figure}

\noindent \textbf{Superconducting Gap Structure from BQPI}\\
After establishing the direction of the gap minima in \fesesonenine{}, we now use the measured $q_7$ dispersion to construct the angular dependence of the gap magnitude, $|\Delta(\theta)|$.
Spatial inhomogeneity and other band features can blur the behavior of $\vec{q_7}$ at small $\vec{q}$. To suppress these effects, we divide the $B=0$\,T dispersion map (Fig.\,\ref{fig:3}b) by its counterpart at $B=3$\,T (Fig.\,\ref{fig:3}d), resulting in a normalized dispersion map, Fig.\,\ref{fig:3}f. The resulting $\vec{q_7}(E)$ dispersion is then geometrically inverted to determine the $|\Delta(\theta)|$, a established technique successfully employed in the determination of the gap structure in other unconventional superconductors \cite{Sprau_Science_orb_selective_cooper, mcelroy2003octet, allan2013imaging}.
As detailed in the Supplementary Note III, we analyze individual constant $\vec{q}$ cuts of these data, shown in Fig.\,\ref{fig:3}g. Each value of $\vec{q_7}$ maps to a value of $\theta$ with a corresponding gap amplitude that is determined by the location of the peaks in the data shown in Fig.\,\ref{fig:3}g. 
The result of this analysis is the extraction of $|\Delta(\theta)|$ from the BQPI data with great precision and over a significantly large range of $\theta$, as shown in Fig.\,\ref{fig:4}a.

\noindent \textbf{Superconducting Gap from Nematic Fluctuations}\\
To deduce the pairing interaction in \fesesonenine, we next compare $|\Delta(\theta)|$ obtained from our experiments to the angular dependence of the gap expected for the two leading pairing candidates: nematic and spin fluctuations. First, we consider the nematic-fluctuations scenario, motivated by the fact that the nematic transition temperature is suppressed to zero at $x=x_c\approx0.17$ in the phase diagram of \feses{}, which coincides with a regime of significantly enhanced nematic fluctuations \cite{hosoi2016nematic}. The distinguishing feature of the electronic interactions mediated by nematic fluctuations is the presence of the nematic form factor $f(\theta) = \lambda \cos(2\theta)$ that describes the angular dependence of the coupling between the electronic and nematic degrees of freedom. This form factor is enforced by the tetragonal symmetry of the lattice and leads to so-called cold spots at $\theta = 45^\circ$ (and symmetry-related points) in any Fermi pocket centered at the $\Gamma$ point. At these cold spots, electrons are nearly decoupled from critical nematic fluctuations, since $f(45^\circ) = 0$. Consequently, if pairing is mediated by nematic fluctuations, the gap should be significantly suppressed at these cold spots, as discussed in Refs.\,\cite{Lederer_enhancement_PRL_2015,Klein-PRB(2018)}. Motivated by these results, we consider the following phenomenological form for the gap function at the central hole pocket: 
\begin{equation}
    \Delta(\theta) = \Delta_s + {\Delta^\prime_s}\cos^2(2\theta).
    \label{Eq:1}
\end{equation}
To test this idea, we fit Eq.\,\ref{Eq:1} to the experimental data, as shown in Fig.\,\ref{fig:4}a,b. We find that Eq.\,\ref{Eq:1} not only accurately describes the data but it also yields a large ratio ${\Delta^\prime_s}/\Delta_s  \approx 8$. This large ratio implies a strong decrease of the gap at the cold spots, and is thus consistent with pairing dominated by nematic fluctuations. 

To go beyond this qualitative description, we follow Ref.\,\cite{Klein-PRB(2018)} and employ an Eliashberg approach to solve a simple model in which pairing on the hole pocket is mediated by quantum critical nematic fluctuations (details in the Supplementary Note V), {without presuming a specific form of $\Delta(\theta)$.} Even though the nematic form factor $f(\theta) = \lambda \cos(2\theta)$ vanishes at the cold spots ($\theta = 45^\circ$ and symmetry-related angles), the temporal fluctuations of the nematic order parameter endow them with a finite gap value, in agreement with the findings of Ref.\,\cite{Klein-PRB(2018)}. Although this low-energy model does not capture the complexity of the band structure of the FeSCs, it has the advantage of having 
{only two parameters -- the distance to the putative QCP $r_0$ and the dimensionless nematic coupling constant $\lambda$. The generic behavior of the gap function that solves the Eliashberg equations consists of deep minima along $45^\circ$, demonstrating that this behavior is a robust feature of the nematic-fluctuations scenario. Importantly, the gap anisotropy is robust for $r_0$ values around $r_0=0$, as we show in Supplementary Note V. This implies that small deviations from the nematic QCP should not lead to large changes in the gap form.  The coupling constant  $\lambda$ controls the depth of the gap minimum and, as a result, the functional form of the gap. As shown in Fig. \ref{fig:4}b (right-hand side of the panel), we find that for a moderate value of the nematic coupling constant, $\lambda = 0.1$, the calculated gap function is in very good agreement with the data and approximately displays the qualitative form proposed in Eq.\,\ref{Eq:1}.} 


\noindent \textbf{Spin Fluctuations Scenarios}\\
Given their primary role in other Fe-based superconductors, it is important to discuss the role of spin fluctuations in the phase diagram of \feses{ }and whether they can give rise to a gap of the form of Eq.\,\ref{Eq:1}. While nematic fluctuations show a diverging behavior at $x_c$ \cite{hosoi2016nematic}, NMR experiments have shown that spin fluctuations are suppressed above $x = 0.09$ \cite{Wiecki_NMR_2018}, {indicating a significant increase in relative strength between nematic and spin fluctuations at the QCP for \fesesonenine.} Recent transport measurements further corroborate the absence of magnetic quantum critical fluctuations in unpressurized \feses{} \cite{Ayres2022transport}. 
{Parallel to this experimental context,} we also find that the gap functions from spin fluctuations, as predicted by theory, are in disagreement with the data. To see this, we note that spin fluctuations in FeSCs are expected to promote a sign-changing gap \cite{fernandes2022iron}, which can be either a $d$-wave or a so-called $s^{\pm}$ gap \cite{Hirschfeld2011,Chubukov2012}. The former emerges when the low-energy spin fluctuations are peaked at the $(\pi,\pi)$ wave vector of the 1-Fe Brillouin zone, whereas the latter appears when the peak is at $(\pi,0)$/$(0,\pi)$. Neutron scattering experiments on undoped FeSe show that $(\pi,0)$/$(0,\pi)$ fluctuations dominate at low energies, and that $(\pi,\pi)$ fluctuations only become relevant at high energies \cite{Wang2016magnetic,Wang2016_neutron, Chen2019anisotropic}, which poses challenges to a $d$-wave pairing scenario.
{Furthermore, a simple $d$-wave function, $\Delta(\theta) = \Delta_d \cos(2\theta)$, does not fit the data (as indicated by the green line in Fig.\,\ref{fig:4}b and Supplementary Note VI), although including higher harmonics could bring small variations of this basic functional form.}

As for the $s^{\pm}$ gap structure generated by $(\pi,0)$/$(0,\pi)$ spin fluctuations, {the predictions, based on various theoretical approaches, invariably result in hole pockets with nearly-isotropic gaps or gaps with minima along the Fe-Fe direction \cite{Graser_2009, Aoki_2009, Maier_2009, Ikeda_2010, Maiti_2011, yin2014spin, Rhodes2021, Fernandez_Martin_RPA}. Importantly, although the near-nodal gap structure observed in our data is never realized in $(\pi,0)$/$(0,\pi)$ spin fluctuations models, it is naturally satisfied in the case of pairing mediated by nematic fluctuations, as discussed earlier.}
We emphasize that our conclusion that nematic fluctuations play a dominant role in promoting pairing in \fesesonenine{ }does not mean that spin fluctuations are irrelevant. In fact, because nematic fluctuations are peaked at zero momentum, whereas spin fluctuations are peaked at large momentum, the latter are likely relevant to determine the relative sign between the gaps on the hole and electron pockets.

\noindent
\textbf{Enhanced low-energy density of states}\\
Our STS measurements also shed some light on the origin of the enhanced DOS near $E=0$ (see Fig.\,\ref{fig:1}e and \cite{Hanaguri_Sci_Adv, mizukami2023unusual}), which has been proposed to emerge from an ultranodal superconducting state with Bogoliubov Fermi surfaces \cite{Agtenberg_PRL_Bogoliubov, setty2020topological, Ultranodal_Review}. 
The gap structure depicted in Fig.\,\ref{fig:4}a,b indicates that a significant portion of momenta have gap values below approximately $150$\,$\mu$eV, comparable to the thermal broadening in the measurements, which naturally leads to a finite tunneling density of states at zero energy.
We also note that recent calculations have found that the superconducting state mediated by quantum critical nematic fluctuations displays enhanced quasi-particle excitations at very low energies due to the nematic cold spots located at $\theta = 45^\circ$ \cite{Chubukov2023}. Such an enhanced DOS at low energies results in a specific heat with a characteristic temperature dependence that agrees well with thermodynamic data in \feses{ }near $x_c$ \cite{mizukami2023unusual}.
{Additionally, in a multiband system, such as \feses{}, different Fermi surface pockets are expected to have gaps with different amplitudes. The \gqe{ }data in the non-superconducting state (Figs.\,\ref{fig:3}d-e) clearly show various additional QPI features at small $|\vec{q}|$ and, although their overlap in $\vec{q}$ space precludes the precise identification of their corresponding initial and final $\vec{k}$ states, those features are also clearly affected by the emergence of superconductivity. The effects of superconductivity on these smaller-$|\vec{q}|$ features are most clearly visible along the $\theta = 0^\circ$ direction (Fig.\,\ref{fig:3}c). 
Interestingly, these inner QPI features lose their intensity at energies within approximately $\pm0.15$\,meV, implying smaller superconducting gaps on the electron $M$ pockets and the inner hole $\Gamma$ pocket.
Although we are not able to determine $\Delta(\theta)$ for these pockets, it is worth noting that determining $\Delta(\theta)$ for the outer hole pocket was already a formidable challenge, as to our knowledge this is the smallest gap for which its angular structure has been determined in an FeSC (Supplementary Note VII). Overall, the combination of such small gaps and pairing mediated by nematic fluctuations, resulting in the $\Delta(\vec{k})$ structure resolved in our measurements, may provide a natural explanation for the observed finite zero-bias DOS. 
}

\noindent \textbf{Conclusions}\\
{Our detailed STS measurements reveal the existence of superconductivity mediated by nematic fluctuations in \fesesonenine. This conclusion is underscored by four key observations:
(\textit{i}) In \fesesonenine{}, which is near a nematic QCP but away from the region of maximum spin fluctuations, we report the first experimental observation of a near nodal gap with minima along $45^\circ$ in an FeSC.
(\textit{ii}) Across all other tetragonal Fe-based superconductors, where nematic and spin-density wave (SDW) phases are tightly bound, experimentally reported gaps \cite{Richard_2015,Allan_LiFeAs} on the hole pockets are nearly isotropic or display shallow minima along the Fe-Fe direction, or both, consistent with pairing attributed to spin fluctuations -- see Fig.\,\ref{fig:4}c.
(\textit{iii}) Theoretical calculations of the spin-mediated $s^\pm$ scenarios predict gaps that are nearly isotropic or exhibit minima along the Fe-Fe direction \cite{Graser_2009, Aoki_2009, Maier_2009, Ikeda_2010, Maiti_2011, yin2014spin, Rhodes2021, Fernandez_Martin_RPA}.
(\textit{iv}) In contrast to predictions based on spin fluctuations, theory distinctly indicates that nematic fluctuations result in a highly anisotropic gap characterized by pronounced (near-nodal) minima at $45^\circ$ from the Fe-Fe direction \cite{Lederer_enhancement_PRL_2015,Klein-PRB(2018)}, exactly as observed in our data.}

Our new finding also puts us in a position to provide a holistic phenomenological description of \feses{}, consistent with various theoretical and experimental studies of superconductivity, nematicity and spin fluctuations, as summarized in Fig.\,\ref{fig:4}d. For small values of $x$ (SC1 region), $(\pi,0)$/$(0,\pi)$ spin fluctuations (described by the magnetic susceptibility $\chi_M$) play a dominant role in the pairing, as indicated by neutron scattering measurements of the spin resonance across $T_c$ \cite{Wang2016magnetic,Wang2016_neutron, Chen2019anisotropic} by NMR studies that correlate the maximum of $T_c$ to the maximum of spin fluctuations near $x=0.09$ \cite{Wiecki_NMR_2018}. 
The strong twofold anisotropy of the gap in \fese{}, manifested as sharp minima along the $x$ or $y$ directions seen by STS \cite{Sprau_Science_orb_selective_cooper} and ARPES \cite{Liu2018}, can be accurately captured by the combined effects, on the pairing state, of long-range nematicity, described by the order parameter $\varphi$, and spin fluctuations \cite{Kang_Fernandes_Chubukov_PRL}. Thus, for $x < x_c$, it is the combination of $\varphi$ and $\chi_M$ that gives rise to a gap with strong minima along the $x$ or $y$ directions, which characterizes the SC1 state. 
On the other hand, across $x_c$, there is no nematic order ($\varphi=0$) and the magnetic susceptibility $\chi_M$ is {relatively} suppressed \cite{Wiecki_NMR_2018}. Conversely, it is the nematic fluctuations (denoted by $\chi_N$) that become large in the vicinity of the QCP, as seen by elasto-resistance measurements \cite{hosoi2016nematic}. Under these conditions, nematic fluctuations dominate the pairing interaction and the gap structure in SC2 becomes strongly anisotropic, with gap minima along $a$ and $b$, as established by our experiments. Therefore, nematicity likely plays an important dual role for the superconductivity in \feses{}, with long-range nematic order inducing gap minima along $x$ and $y$ but with nematic fluctuations inducing near nodes along $a$ and $b$. 
Altogether, our work establishes a new avenue for studying the relationship between nematic fluctuations and superconductivity in other materials.

\begin{acknowledgments}
\noindent We thank Andrey Chubukov, Leonid Glazman and Pavlo Sukhachov for fruitful discussions during the preparation of this manuscript. 
E.H.d.S.N acknowledges support by the National Science Foundation under Grant No.\,DMR-2034345. This work was supported by the Alfred P. Sloan Fellowship (E.H.d.S.N.). Sample synthesis was supported by the UC Lab Fees Research Program (LFR-20-653926). R.M.F. (theory work) was supported by the U.S. Department of Energy, Office of Science, Basic Energy Sciences, Materials Science and Engineering Division, under Award No. DE-SC0020045. E.M. acknowledges support by CNPq-Brazil, grant No. 309584/2021-3.

\noindent\textbf{Author contributions:} 
P.K.N., K.S., M.W. and E.H.D.S.N.
performed the STM measurements with the assistance of X.Y., A.G.G., and A.G..
V.S.d.C, R.M.F. and E.M. performed theoretical calculations. X.Y. computed QPI simulations with assistance from A.G. and E.H.D.S.N.. J.K.B. grew and characterized the \feses{ }crystals with support from P.K. and V.T..  E.H.D.S.N., R.M.F., P.K.N., K.S., V.S.d.C, A.G. and E.M. wrote the manuscript with input from all other authors. E.H.d.S.N. conceived of the experiments and was responsible for overall project direction, planning, and management. 



\end{acknowledgments}

\section*{
REFERENCES
}

\bibliographystyle{eh_style}
\bibliography{bib_lib}

\end{document}